\documentclass[reprint,
superscriptaddress,amsmath,amssymb,aps, longbibliography
]{revtex4-2}

\usepackage{graphicx}
\usepackage{dcolumn}
\usepackage{bm}
\usepackage{siunitx}
\usepackage[noend]{algorithmic}
\usepackage{algorithm}
\usepackage[skins]{tcolorbox}
\usepackage{setspace}
\usepackage{blkarray}

\usepackage{xcolor}
\usepackage{soul}
\usepackage{systeme}

\usepackage{scalerel} 

\makeatletter
\def\maketag@@@#1{\hbox{\m@th\normalfont\normalsize#1}}
\makeatother

\newcommand{\me}{\mathrm{e}}
\newcommand{\sgn}{\text{sgn}}

\usepackage{booktabs}
\usepackage{hyperref}
\usepackage[capitalize]{cleveref}
\usepackage[utf8]{inputenc}

\usepackage[lining,semibold]{libertine} 
\usepackage{amsthm}
\usepackage[libertine, cmintegrals, bigdelims, vvarbb]{newtxmath}

\usepackage{amsmath}
\usepackage{amsfonts}
\usepackage{mathrsfs}
\usepackage{gensymb}
\usepackage{bbm}
\usepackage{dsfont}

\usepackage{kbordermatrix}
\renewcommand{\kbldelim}{(}
\renewcommand{\kbrdelim}{)}

\usepackage{chemformula}
\usepackage[caption=false]{subfig}

\usepackage{soul}
\usepackage{xcolor}

\theoremstyle{definition}

\usepackage{scalerel} 

\definecolor{webgreen}{rgb}{0,.5,0}
\definecolor{webbrown}{rgb}{.6,0,0}
\definecolor{grigio}{rgb}{.85,.85,.85} 
\definecolor{RoyalBlue}{rgb}{0.0, 0.14, 0.4}
\definecolor{skyblue1}{rgb}{0.45,0.62,0.81}
\definecolor{skyblue2}{rgb}{0.2,0.39,0.64}
\definecolor{skyblue3}{rgb}{0.13,0.29,0.53}
\definecolor{scarlet1}{rgb}{0.93,0.16,0.16}
\definecolor{scarlet2}{rgb}{0.8,0,0}
\definecolor{scarlet3}{rgb}{0.64,0,0}

\definecolor{g}{gray}{0.50}

\hypersetup{%
    colorlinks=true, linktocpage=true, pdfstartpage=1, pdfstartview=FitV,%
    breaklinks=true, pdfpagemode=UseNone, pageanchor=true, pdfpagemode=UseOutlines,%
    plainpages=false, bookmarksnumbered, bookmarksopen=true, bookmarksopenlevel=1,%
    hypertexnames=true, pdfhighlight=/O,
    urlcolor=webbrown, linkcolor=RoyalBlue, citecolor=webgreen, 
    pdftitle={},%
    pdfauthor={Timur Aslyamov},%
    pdfsubject={},%
    pdfkeywords={},%
    pdfcreator={pdfLaTeX},%
    pdfproducer={LaTeX REVTeX}%
}

\begin{document}
\title{Nonequilibrium Response for Markov Jump Processes: Exact Results and Tight Bounds}

\author{Timur Aslyamov}
\email{timur.aslyamov@uni.lu}

\author{Massimiliano Esposito}
\email{massimiliano.esposito@uni.lu}
\affiliation{Department of Physics and Materials
Science, University of Luxembourg, L-1511 Luxembourg City, Luxembourg}

\date{\today}

\begin{abstract}
Generalizing response theory of open systems far from equilibrium is a central quest of nonequilibrium statistical physics. Using stochastic thermodynamics, we develop an algebraic method to study the response of nonequilibrium steady state to arbitrary perturbations. This allows us to derive explicit expressions for the response of edge currents as well as traffic to perturbations in kinetic barriers and driving forces. We also show that these responses satisfy very simple bounds. For the response to energy perturbations, we straightforwardly recover results obtained using nontrivial graph-theoretical methods.
\end{abstract}
\maketitle


\textit{Introduction.}---Linear response theory is a central tenet in statistical physics \cite{marconi2008fluctuation,polettini2019effective,owen2023size}.
The response of systems in steady state at, or close to, equilibrium is described by the seminal dissipation-fluctuation relation (DFR) \cite{kubo1966fluctuation,forastiere2022linear}. 
Generalizations to study the response of systems in nonequilibrium steady state (NESS) are a more recent endeavor, in particular since the advent of stochastic thermodynamics \cite{agarwal1972fluctuation,seifert2010fluctuation,prost2009generalized,altaner2016fluctuation,baiesi2009fluctuations,baiesi2013update,baldovin2022many,hatano2001steady,falasco2019negative,shiraishi2022time,dechant2020fluctuation}. Understanding the response of far-from-equilibrium systems is of great conceptual but also practical importance (e.g. to characterize homeostasis, design resilient nanotechnologies, detect critical transitions, and metabolic control).
Progress in this direction relies on our ability to derive useful expressions for NESS responses, and possibly derive practically meaningful bounds for them. 

In this Letter, we study the NESS response of Markov jump processes within stochastic thermodynamics. In this context, Ref.~\cite{owen2020universal} constitute a frontier. The authors studied the response to perturbations of the energy landscape parameters. 
They derived an exact result and two bounds using graph-theoretic methods, which can be quite tedious and nonintuitive to use~\cite{king1956schematic,hill1966studies,schnakenberg1976network}. 
We develop a novel approach based on simple linear algebra, which allows us to go significantly further than currently known results. 
We first derive a simple and elegant expression for the response of a NESS to arbitrary perturbations. We use it to straightforwardly reproduce the main result~\cite{owen2020universal} for the NESS response to perturbations of the energy landscape. But more importantly we also use it to derive novel and simple expressions for the response of edge currents and traffic to kinetic barriers and driving forces perturbations.
We furthermore derive four remarkably simple bounds for these four quantities (see \cref{tab:main-results}), which can be added to the list of simple bounds valid far-from-equilibrium, together with thermodynamic uncertainty relations \cite{barato2015thermodynamic,gingrich2016dissipation} and speed limits \cite{falasco2020dissipation}. 

\textit{Setup.}---We consider a Markov jump process over a discrete set of $N$ states. Transitions between these states are described by the rate matrix $\mathbb{W}/\tau$, where the element $W_{nm}/\tau$ defines the probability per unit time $\tau$ to jump from state $m$ to state $n$. Below we choose $\tau=1$, to adimensionalize the matrix $\mathbb{W}$.  We assume that all transitions are reversible and that the matrix $\mathbb{W}$ is irreducible \cite{van1992stochastic}.
This ensures the existence of a unique steady-state probability distribution $\boldsymbol{\pi}=(\pi_1,\dots,\pi_N)^\intercal$ satisfying 
    \begin{align}\label{eq:master_eq}
        \mathbb{W}\cdot\boldsymbol{\pi}&=\boldsymbol{0}\,,
    \end{align}
where the length of the vector $\boldsymbol{\pi}$ is $|\boldsymbol{\pi}|=1$.
When the rates depend on a model parameter $\eta$, one can define the linear response (resp. sensitivity) of the nonequilibrium state as $\partial_\eta q$ (resp. $\partial_\eta\ln q$) for an arbitrary quantity $q$. 

\textit{General theory.}---The rate matrix $\mathbb{W}$ in \cref{eq:master_eq} has only one zero eigenvalue \cite{van1992stochastic}. This allows us to rewrite \cref{eq:master_eq} as
\begin{subequations}
\begin{align}
\label{eq:master_eq_K}
&\mathbb{K}_n\cdot\boldsymbol{\pi} =\boldsymbol{e}_n\,,\\
\label{eq:matK}
&\mathbb{K}_n=
\begin{blockarray}{cccccc}
\begin{block}{cc (cccc)}
\color{gray} 1 & & W_{11} & W_{12} & \dots & W_{1N} \\
\color{gray} \vdots & & \vdots & \vdots & \dots & \vdots  \\
\color{gray} n-1 & & W_{n-1,1} & W_{n-1,2} & \dots & W_{n-1,N} \\
\color{gray} n & & 1 & 1 & \dots & 1  \\
\color{gray} n+1 & & W_{n+1,1} & W_{n+1,2} & \dots & W_{n+1,N} \\
\color{gray} \vdots & & \vdots & \vdots & \dots & \vdots  \\
\color{gray} N & & W_{N1} & W_{N,2} & \dots & W_{N,N} \\
\end{block}
\end{blockarray}\,\,,
\end{align}
\end{subequations}
where $\boldsymbol{e}_n$ denotes the vector with a $1$ for the $n$-th element and $0$'s elsewhere, and where the matrix $\mathbb{K}_n$ coincides with the rate-matrix $\mathbb{W}$ except the $n$-th row. 
Since the matrix $\mathbb{K}_n$ is invertible [$\det \mathbb{K}_n\neq 0$] the solution of \cref{eq:master_eq_K} has the following form:
\begin{equation}
\label{eq:ss}
    \boldsymbol{\pi}=\mathbb{K}_n^{-1}\cdot\boldsymbol{e}_n\,.
\end{equation}
To find the linear response $\partial_\eta\boldsymbol{\pi}$ we calculate the derivative $\partial_\eta$ of \cref{eq:master_eq_K}:
\begin{align}
\label{eq:master_eq_K_perturb}
    \partial_\eta [\mathbb{K}_n(\eta)&\cdot\boldsymbol{\pi}(\eta)] =\boldsymbol{0}\,, \nonumber\\
    \mathbb{K}_n\cdot\partial_\eta\boldsymbol{\pi} & = - \partial_\eta \mathbb{K}_n\cdot\boldsymbol{\pi}\,.
\end{align}
Solving \cref{eq:master_eq_K_perturb}, we arrive at the desired result:
\begin{align}
\label{eq:response}
    \partial_\eta\boldsymbol{\pi} &=-\mathbb{K}_n^{-1}\cdot\partial_\eta \mathbb{K}_n\cdot\boldsymbol{\pi}\,.
\end{align}
\Cref{eq:response} will be central in what follows. Indeed, it provides a linear algebra-based method to calculate different nonequilibrium responses which is much simpler and direct than methods based on graph theory representations of $\boldsymbol{\pi}$ \cite{owen2020universal}. At this stage, \cref{eq:response} holds for any dependence of $\mathbb{W}(\eta)$ on the control parameter.

\textit{Rate-matrix model.}---To proceed, we follow Ref.~\cite{owen2020universal} and parameterize the nondiagonal elements of the rate matrix as
\begin{equation}
    \label{eq:W}
    W_{ij}=\me^{-(B_{ij}-E_j-F_{ij}/2)}\,,
\end{equation}
where $E_j$ are the vertex parameters, $B_{ij}=B_{ji}$ are the symmetric edge parameters, and $F_{ij}=-F_{ji}$ are the antisymmetric edge parameters. Expression \eqref{eq:W} is reminiscent of Arrhenius rates that characterize the transition rates of a system in an energy landscape with wells of depths $E_j$, connected via barriers of heights $B_{ij}$, and subjected to nonconservative driving forces $F_{ij}$ along the transition paths~\cite{owen2020universal,falasco2021local}. 
These rates satisfy local detailed balance ensuring the compatibility with stochastic thermodynamics \cite{rao2018conservation,maes2021local,falasco2021local}.

\textit{Vertex parameters.}---To calculate $\partial_{E_{n}}\boldsymbol{\pi}$, we note that only the $n$-th column of the matrix $\mathbb{K}_n$ depends on $E_{n}$. Therefore, 
\begin{align}
\label{eq:dKdE}
\partial_{E_n}\mathbb{K}_n &=
\begin{blockarray}{ccccccccc}
& & \color{gray} \phantom{1}  & \color{gray} \phantom{0} & \color{gray}\phantom{n-1} & \color{gray} n & \color{gray} \phantom{n+1} & \color{gray}\phantom{0} & \color{gray} \phantom{N} & \\
\begin{block}{cc (ccccccc)}
\color{gray} 1 & & \phantom{0} & \phantom{0} & \phantom{0} & W_{1,n} & \phantom{0} & \phantom{0} & \phantom{0} \\
\color{gray}\vdots   & & \phantom{0} & \phantom{0} &\phantom{0} & \vdots &\phantom{0} & \phantom{0} & \phantom{0} \\
\color{gray} n-1 & & \phantom{0} & \phantom{0} & \phantom{0} & W_{n-1,n} & \phantom{0} & \phantom{0} & \phantom{0} \\
\color{gray} n & & \phantom{0} & \phantom{0} & \phantom{0} & 0 & \phantom{0} & \phantom{0} & \phantom{0} \\
\color{gray} n+1 & & \phantom{0} & \phantom{0} & \phantom{0} & W_{n+1,n} & \phantom{0} & \phantom{0} & \phantom{0} \\
\color{gray} \vdots  & & \phantom{0} & \phantom{0} &\phantom{0} & \vdots &\phantom{0} & \phantom{0} & \phantom{0} \\
\color{gray} N & & \phantom{0} & \phantom{0} & \phantom{0} & W_{N,n} & \phantom{0} & \phantom{0} & \phantom{0} \\
\end{block}
\end{blockarray}\,\,,
\end{align}
where all columns but the $n$-th one are zero. The element $(n,n)$ is zero because $K_{n,n}=1$. Here and below, empty spaces in matrices denote zeros. 
Inserting \cref{eq:dKdE} into \cref{eq:response}, we immediately recover a key result of Ref.~\cite{owen2020universal} obtained using nontrivial graph-theoretical methods, namely 
\begin{align}
\label{eq:response_En}
    \partial_{E_n}\boldsymbol{\pi} &=-\pi_n\mathbb{K}_n^{-1}\cdot (\boldsymbol{K}_n-\boldsymbol{e}_n)
    =-\pi_n(\boldsymbol{e}_n-\boldsymbol{\pi})\,,
\end{align}
where $\boldsymbol{K}_n$ is the $n$-th column of $\mathbb{K}_n$ and where we used \cref{eq:ss}. 

\textit{Symmetric edge parameters.}---We proceed with calculating $\partial_{B_{nm}}\boldsymbol{\pi}$. One can see from \cref{eq:W} that such a perturbation changes $W_{nm}$ and $W_{mn}$. These rates are also contained in the diagonal elements of the matrix $\mathbb{W}$ since $W_{nn}=-\sum_{m\neq n}W_{mn}$. Overall four elements depend on $B_{nm}$: $W_{nm}$, $W_{nn}$, $W_{mn}$, $W_{mm}$. But the matrix $\mathbb{K}_n$ defined in \cref{eq:matK} only contains two of those elements (due to row $n$). Using \cref{eq:W}, their derivatives reads $\partial_{B_{nm}}W_{mn}=- W_{mn}$ and $ \partial_{B_{nm}}W_{mm}=- \partial_{B_{nm}}W_{nm}= W_{nm}$, and we find that
\begin{align}
\label{eq:dKdB}
\partial_{B_{nm}}\mathbb{K}_n&=
\begin{blockarray}{ccccccccc}
& & \color{gray} 1 & \color{gray} \dots & \color{gray}n & \color{gray} \dots & \color{gray} m & \color{gray}\dots & \color{gray}N & \\
\begin{block}{cc (ccccccc)}
\color{gray} 1 & & \phantom{0} & \phantom{0} & \phantom{0} & \phantom{0} & \phantom{0} & \phantom{0} & \phantom{0} \\
\color{gray} \vdots & & \phantom{0} & \phantom{0} & \phantom{0} & \phantom{0} & \phantom{0} & \phantom{0} & \phantom{0} \\
\color{gray} m & & \phantom{0} & \phantom{0} & -W_{mn} & \phantom{0} & W_{nm} & \phantom{0} & \phantom{0} \\
\color{gray} \vdots & & \phantom{0} & \phantom{0} & \phantom{0} & \phantom{0} & \phantom{0} & \phantom{0} & \phantom{0}\\
\color{gray} N & & \phantom{0} & \phantom{0} & \phantom{0} & \phantom{0} & \phantom{0} & \phantom{0} & \phantom{0}\\
\end{block}
\end{blockarray}\,\,.
\end{align}
Calculating $(\partial_{B_{nm}}\mathbb{K}_n)\cdot\boldsymbol{\pi}$ and inserting into \cref{eq:response} we get:
     \begin{align}
      \label{eq:response_Bnm}
        \partial_{B_{nm}}\boldsymbol{\pi}& = -\mathbb{K}_n^{-1}\cdot \boldsymbol{e}_m (W_{nm}\pi_m-W_{mn}\pi_n) = -\frac{\boldsymbol{\kappa}}{\det \mathbb{K}_n}J_{nm} \,,
     \end{align}
where we recognize the NESS current $J_{nm}=W_{nm}\pi_m-W_{mn}\pi_n$ from $m$ to $n$, and where $\boldsymbol{\kappa}/\det\mathbb{K}_n$ is the $m$-th column of the matrix $\mathbb{K}_n^{-1}$. The elements $\kappa_i$ can be defined in terms of the minors $M_{im}(\mathbb{K}_n^\intercal)$ of the matrix $\mathbb{K}_n^\intercal$:
\begin{equation}
\label{eq:kappa}
    \kappa_i=(-1)^{i+m}M_{im}(\mathbb{K}_n^\intercal)=(-1)^{i+m}M_{mi}(\mathbb{K}_n)\,.
\end{equation}
Since the minors $M_{mi}(\mathbb{K}_n)$ do not include the $m$-th row of the matrix $\mathbb{K}_n$, the elements $\kappa_i$ do not depend on $B_{nm}$ and $F_{nm}$ [see \cref{eq:matK}]. 

Expression \eqref{eq:response_Bnm} is a new result. In Ref.~\cite{owen2020universal}, only the following bound was obtained: $|\partial_{B_{nm}}\pi_i|\leq \pi_i(1-\pi_i)\tanh(F_\text{max}/4)$, where $F_\text{max}$ is the maximum absolute value of the affinity along all cycles containing the edge $(n,m)$. A numerical comparison between the two is given in \cref{sec:example}, see \cref{fig:exact}. 
A direct implication of our result is that the response is suppressed, $\partial_{B_{nm}}\boldsymbol{\pi}=0$, when the edge $(n,m)$ is detailed balanced $J_{nm}=0$. Instead, ensuring the suppression of the response from the bound~\cite{owen2020universal}, implies the more restrictive condition $F_\text{max}=0$, which corresponds to equilibrium where all edge currents vanish. An example where an edge current vanishes while the forces are non-zero is provided in \cref{sec:example}. They have also been shown to produce ``Green-Kubo-like'' FDR~\cite{altaner2016fluctuation}.

\textit{Antisymmetric edge parameters.}---We now calculate $\partial_{F_{nm}}\boldsymbol{\pi}$ from \Cref{eq:response}. The non-zero elements of $\partial_{F_{nm}}\mathbb{K}_n$ are $\partial_{F_{nm}}W_{mn}=-W_{mn}/2$ and $\partial_{F_{nm}}W_{mm}=-\partial_{F_{nm}}W_{nm}=-W_{nm}/2$: 
\begin{align}
\label{eq:dKdF}
\partial_{F_{nm}}\mathbb{K}_n&=
\begin{blockarray}{ccccccccc}
& & \color{gray} 1 & \color{gray} \dots & \color{gray}n & \color{gray} \dots & \color{gray} m & \color{gray}\dots & \color{gray}N & \\
\begin{block}{cc (ccccccc)}
\color{gray} 1 & & \phantom{0} & \phantom{0} & \phantom{0} & \phantom{0} & \phantom{0} & \phantom{0} & \phantom{0} \\
\color{gray} \vdots & & \phantom{0} & \phantom{0} & \phantom{0} & \phantom{0} & \phantom{0} & \phantom{0} & \phantom{0} \\
\color{gray} m & & \phantom{0} & \phantom{0} & -W_{mn}/2 & \phantom{0} & -W_{nm}/2 & \phantom{0} & \phantom{0} \\
\color{gray} \vdots & & \phantom{0} & \phantom{0} & \phantom{0} & \phantom{0} & \phantom{0} & \phantom{0} & \phantom{0}\\
\color{gray} N & & \phantom{0} & \phantom{0} & \phantom{0} & \phantom{0} & \phantom{0} & \phantom{0} & \phantom{0}\\
\end{block}
\end{blockarray}\,\,.
\end{align}
Using \cref{eq:response}, we arrive at:
\begin{align}
\label{eq:response_Fnm}
    \partial_{F_{nm}}\boldsymbol{\pi}& 
    = \mathbb{K}_n^{-1}\cdot \boldsymbol{e}_m \frac{W_{nm}\pi_m+W_{mn}\pi_n}{2}
    = \frac{\boldsymbol{\kappa}}{\det\mathbb{K}_n}\frac{\tau_{nm}}{2}\,,
\end{align}
where $\tau_{nm}=W_{nm}\pi_m+W_{mn}\pi_n$ is the edge traffic (related to the expected escape rate, activity and frenesy~\cite{maes2020frenesy}).
In Ref.~\cite{owen2020universal}, only the following bound was obtained $|\partial_{F_{nm}}\pi_i|\leq \pi_i(1-\pi_i)$. 

\textit{Responses of current and traffic.}---Using \cref{eq:response_Bnm,eq:response_Fnm}, the sensitivities of the edge currents reads:
    \begin{subequations}
    \label{eq:J-response}
    \begin{align}
        \label{eq:Jnm-response-Bnm}
        \partial_{B_{nm}}\ln J_{nm} &= -1+\Delta_{nm}\,, \\
        \label{eq:Jnm-response-Fnm}
       \partial_{F_{nm}}\ln J_{nm} &=\frac{\tau_{nm}}{J_{nm}}\frac{(1-\Delta_{nm})}{2}\,,\\
        \label{eq:Delta}
        \Delta_{nm} &= \frac{W_{mn}\kappa_n-W_{nm}\kappa_m}{\det\mathbb{K}_n}\,.
    \end{align}      
    \end{subequations}
Similarly, the sensitivities of edge traffic reads: 
\begin{subequations}
\label{eq:tau-response}
    \begin{align}
        \label{eq:taunm-response_Bnm}
        \partial_{B_{nm}}\ln\tau_{nm} &= -1-\frac{J_{nm}}{\tau_{nm}}\nabla_{nm}\,, \\
        \label{eq:taunm-response-Fnm}
        \partial_{F_{nm}} \ln \tau_{nm} &=\frac{1}{2}\bigg(\frac{J_{nm}}{\tau_{nm}}+\nabla_{nm}\bigg)\,,\\
        \label{eq:nabla}
        \nabla_{nm} &= \frac{W_{mn}\kappa_n+W_{nm}\kappa_m}{\det\mathbb{K}_n}\,.
    \end{align}
\end{subequations}
These two results, \cref{eq:J-response,eq:tau-response}, are important because they provide explicit algebraic expressions for the response. Indeed, the variables $\Delta_{nm}$ and $\nabla_{nm}$ defined in \cref{eq:Delta,eq:nabla} do not depend on $\pi_i$. They depend only on the elements in the minors $(m,n)$ and $(m,m)$ of the matrix $\mathbb{K}_n$.

\begin{table}[h!]
    \centering
    \includegraphics[width=0.48\textwidth]{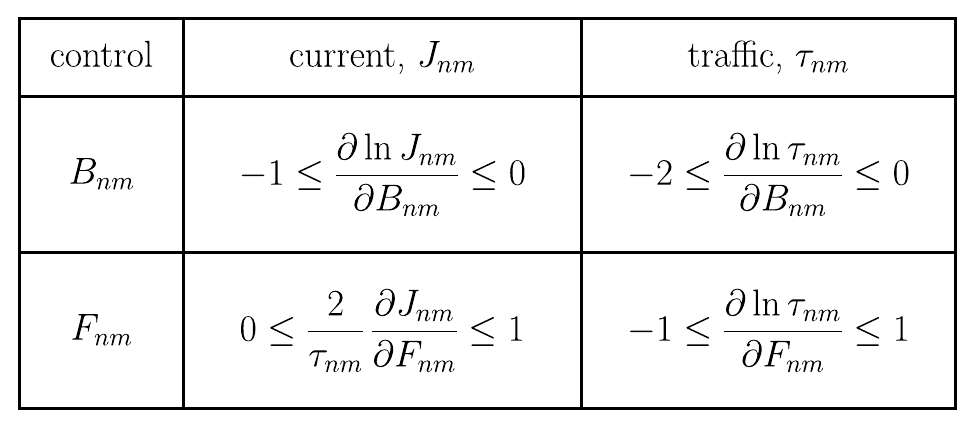}
    \caption{The central and right columns correspond to the response of the current and traffic, respectively. The central and bottom rows are perturbations of the symmetric and antisymmetric edge parameters, respectively.}
    \label{tab:main-results}
\end{table}
\textit{Bounds and discussion}.---Another important result is that simple bounds can be obtained for \cref{eq:J-response,eq:tau-response}. They are given in \cref{tab:main-results} and bound the sensitivities $\partial_\eta \ln q$ for all combinations of $q=\{J_{nm},\tau_{nm}\}$ and $\eta=\{B_{nm},F_{nm}\}$. In \cref{sec:proof}, we derive the following bounds for $\Delta_{nm}$ and $\nabla_{nm}$,
\begin{subequations}
    \label{eq:Delta-nabla-bounds}
    \begin{align}
    \label{eq:Delta-bound}
    0 \leq & \Delta_{nm}\leq 1\,,\\
    \label{eq:nabla-bound}
    |\nabla_{nm}|\leq &\Delta_{nm} \leq 1\,,
\end{align}
\end{subequations}
which can be used to prove all bounds in \cref{tab:main-results}. Indeed, inserting \cref{eq:Delta-bound} into \cref{eq:Jnm-response-Bnm,eq:Jnm-response-Fnm} we get two tight bounds for the current $J_{nm}$ in \cref{tab:main-results}. Using \cref{eq:nabla-bound,eq:taunm-response_Bnm,eq:taunm-response-Fnm}, we derive two tight bounds for the traffic
\begin{subequations}
\label{eq:tau-tight-bounds}
\begin{align}
\bigg|\frac{\tau_{nm}}{J_{nm}}\Big(\frac{\partial \ln \tau_{nm}}{\partial B_{nm}}+1\Big)\bigg|&\leq 1\,, \\
\bigg|\frac{2\partial \ln \tau_{nm}}{\partial F_{nm}}-\frac{J_{nm}}{\tau_{nm}}\bigg|&\leq 1\,.
\end{align}
\end{subequations}
The simpler bounds for $\tau_{nm}$ shown in \cref{tab:main-results} are not tight anymore. They are obtained using \cref{eq:nabla-bound} and $|J_{nm}/\tau_{nm}|\leq 1$ in \cref{eq:taunm-response_Bnm,eq:taunm-response-Fnm}.
To discuss the saturation of the tight bounds in \cref{tab:main-results,eq:tau-tight-bounds}, we consider one of them:
\begin{align}
    \label{eq:Jnm-bound}
    -1 \leq \partial_{B_{nm}}\ln J_{nm}\leq 0\,.
\end{align}
The upper bound in \cref{eq:Jnm-bound} is simple to understand: a higher energy barrier ($B_{nm}$) always results in a lower absolute value of the current between states $n$ and $m$. This bound is saturated at $\Delta_{nm}=1$, which reveals another (topological) way to reduce the response of the current. To saturate the lower bound in \cref{eq:Jnm-bound}, one needs $\Delta_{nm}=0$. However, in \cref{sec:proof} we prove that $\Delta_{nm}=0$ only if $W_{nm}=W_{mn}=0$ or $\kappa_{m}=\kappa_{n}=0$, where the former condition is equivalent to $J_{nm}=0$. Therefore, excluding the case $\kappa_{m}=\kappa_{n}=0$, 
the lower bound of \cref{eq:Jnm-bound} can be saturated only for a zero current. This is illustrated by the numerical simulations shown in \cref{fig:random}, where the set of possible values $\partial_{B_{nm}}\ln J_{nm}$ touches the edge $-1$ only at one point $J_{nm}=0$. 
The bound for the sensitivity $\partial_{F_{nm}}\ln J_{nm}$ has the same properties as \cref{eq:Jnm-bound}. The bounds in \cref{eq:tau-tight-bounds} saturate at $\nabla_{nm}=\pm 1$, which implies $\Delta_{nm}=1$, see \cref{eq:nabla-bound}.

\textit{Future studies.}---Our first main result \cref{eq:response} provides a general algebraic expression of a NESS response with respect to any parameterization of the rate matrix. 
Our other results rely on the Arrhenius-like form \eqref{eq:W} of the rates, which allows us to perturb isolated edges. But our methodology can be extended to consider more general rate matrices with nonconservative force acting on multiple edges \cite{rao2018conservation,falasco2023macroscopic,dechant2020fluctuation}. 
It could also be used to study the stationary responses of other physical observables (beyond currents and activities) \cite{martins2023topologically,chun2023trade}, as well as to study time-dependent ``Green-Kubo-Agarwal-like'' relations \cite{seifert2010fluctuation,baiesi2013update}. 

\textit{Acknowledgments.}---This research was funded by project ChemComplex (C21/MS/16356329). We thank Massimo Bilancioni for detailed feedback on our manuscript.

\appendix
\section{Proof of bounds in \cref{eq:Delta-nabla-bounds}}
\label{sec:proof}
We prove the bounds in \cref{eq:Delta-nabla-bounds}. The determinant $\det \mathbb{K}_n$ on the $m$-th row of the matrix $\mathbb{K}_n$ can be written as
\begin{align}
\label{eq:det-kappa}
    \det \mathbb{K}_n &= (-1)^{m+n}W_{mn}M_{mn}(\mathbb{K}_n)+(-1)^{m+m}W_{mm}M_{mm}(\mathbb{K}_n)\nonumber\\
    &+\sum_{i\neq m,n}(-1)^{i+m}W_{mi}M_{mi}(\mathbb{K}_n)\, \nonumber \\
    &= W_{mn}\kappa_n - W_{nm}\kappa_m + C\,,
\end{align}
where we used \cref{eq:kappa} and where $C$ denotes the sum of all terms which do not depend on $B_{nm}$ and $F_{nm}$. Since $\det\mathbb{K}_n=\prod_{i}^{N-1}\lambda_i$, where $\lambda_i$ are nonzero negative eigenvalues of the matrix $\mathbb{W}$ (see \footnote{We use Theorem 2 from page 157 of Ref.~\cite{lancaster1985theory} to write 
$\det\mathbb{K}_n=(-1)^{N-1}\lim_{\lambda\to 0}\det(\lambda \mathbb{E}-\mathbb{W})/\lambda=(-1)^{N-1}\prod_{i=1}^{N-1}(-\lambda_i)$.}), 
the sign of the determinant $\sgn\det\mathbb{K}_n=(-1)^{N-1}$ is fixed and does not depend on $B_{nm}$ and $F_{nm}$. 
Using the fact that $C$ does not depend on $B_{nm}$, we can determine the sign of $C$ from \cref{eq:det-kappa} in the limit $B_{nm}\to\infty$, where $W_{nm},W_{mn}\to 0$:
\begin{equation}
\label{eq:sgn-C}
    \sgn~C = \lim_{B_{nm}\to\infty}\sgn\det\mathbb{K}_n=(-1)^{N-1}\,.
\end{equation}
Using the fact that the signs of $C$ and $\det{\mathbb{K}_n}$ are the same, we can rewrite \cref{eq:Delta} using \cref{eq:det-kappa} as
\begin{align}
    \label{eq:Delta-exact}
    \Delta_{nm}=1-\Big|\frac{C}{\det\mathbb{K}_n}\Big|\,,
\end{align}
which gives us the upper bound in \cref{eq:Delta-bound}.

In the case $\kappa_n=0$, $\kappa_m=0$, \cref{eq:Delta,eq:nabla} satisfy the bounds \eqref{eq:Delta-bound} and \eqref{eq:nabla-bound}. Considering $\kappa_n\neq0$ and $\kappa_m\neq 0$, the following limits of \cref{eq:det-kappa} hold:
\begin{subequations}
    \begin{align}
        \label{eq:det-limit-B}
        \lim_{B_{nm}\to -\infty}\det\mathbb{K}_n&=W_{mn}\kappa_n - W_{nm}\kappa_m\,,\\
        \label{eq:det-limit-F-inf-1}
        \lim_{F_{nm}\to \infty}\det\mathbb{K}_n&= - W_{nm}\kappa_m\,,\\
        \label{eq:det-limit-F-inf-2}
        \lim_{F_{nm}\to -\infty}\det\mathbb{K}_n&= W_{mn}\kappa_n\,.
    \end{align}
\end{subequations}
Since $\sgn(W_{mn}\kappa_n/\det{\mathbb{K}_n})$ and $\sgn(W_{nm}\kappa_m/\det{\mathbb{K}_n})$ are fixed, we can find them using \cref{eq:det-limit-F-inf-1,eq:det-limit-F-inf-2}
\begin{subequations}
\label{eq:Delta-terms-signs}
    \begin{align}
    \label{eq:Delta-terms-signs-n}
        \sgn\bigg(\frac{W_{mn}\kappa_n}{\det{\mathbb{K}_n}}\bigg)&=\lim_{F_{nm}\to -\infty}\frac{W_{mn}\kappa_n}{\det{\mathbb{K}_n}}=1\,,\\
     \label{eq:Delta-terms-signs-m}  
        \sgn\bigg(\frac{W_{nm}\kappa_m}
        {\det{\mathbb{K}_n}}\bigg)&=\lim_{F_{nm}\to \infty}\frac{W_{nm}\kappa_m}{\det{\mathbb{K}_n}}=-1\,,
    \end{align}
\end{subequations}
which implies the lower bound in \cref{eq:Delta-bound} and 
\begin{align}
    \label{eq:delta-terms-ratio}
    \kappa_n\kappa_m&<0\,.
\end{align}
Combining \cref{eq:Delta-exact,eq:Delta-terms-signs}, we derive the inequalities \eqref{eq:Delta-bound}.

The lower bound in \cref{eq:Delta-bound} is saturated only when $W_{nm}=W_{mn}=0$, while for $W_{nm}\neq0$ the condition in \cref{eq:delta-terms-ratio} implies $\Delta_{nm}\neq 0$. The upper bound in \cref{eq:Delta-bound} is saturated in the limit $B_{nm}\to-\infty$ [see \cref{eq:det-limit-B,eq:Delta}], as well as when $C=0$.

To find bounds for $\nabla_{nm}$, we rewrite it as follows:
\begin{align}
     \label{eq:nabla-delta}
        \nabla_{nm} &=\Delta_{nm} \frac{W_{mn}\kappa_n+W_{nm}\kappa_m}{W_{mn}\kappa_n - W_{nm}\kappa_m}\,.
\end{align}
If $W_{mn}\kappa_n=0$, then $\nabla_{nm}=-\Delta_{nm}$, otherwise we have:
\begin{align}
     \label{eq:nabla-delta-2}
        \nabla_{nm} &=\Delta_{nm} \frac{1+a}{1-a}\,,~\text{where}~~
        a =\frac{W_{nm}\kappa_m}{W_{mn}\kappa_n}\leq 0\,.
\end{align}
Since $|(1+a)/(1-a)|\leq 1$ for $a\leq 0$, we find \cref{eq:nabla-bound}. 

In the case $\kappa_n=0$, $\kappa_m\neq 0$ (resp. $\kappa_n\neq0$, $\kappa_m=0$), we derive \cref{eq:Delta-bound} using \cref{eq:Delta-exact} and \cref{eq:Delta-terms-signs-m} (resp. \cref{eq:Delta-terms-signs-n}); and we have $\nabla_{nm}=- \Delta_{nm}$ (resp. $\nabla_{nm}=\Delta_{nm}$).

\section{Example of network}\label{sec:example}

\begin{figure}[h]
    \centering
    \includegraphics[width=0.45\textwidth]{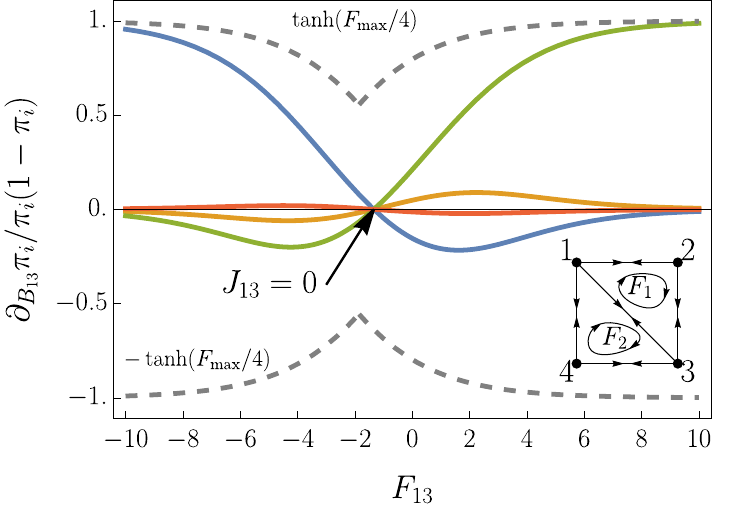}
    \caption{Inset: Example of the network with 4 states; $F_1$ and $F_2$ denote the forces in the cycles $1-2-3-1$ and $1-4-3-1$, respectively. Main: The solid curves show the responses $\partial_{B_{13}\pi_i}$ from \cref{eq:response_Bnm} scaled to $\pi_i(1-\pi_i)$, where $i=1,2,3,4$ correspond to blue, orange, green,  and red colors, respectively. In these coordinates, the dashed lines ($\pm \tanh(F_\text{max}/4)$) correspond to the bound from~\cite{owen2020universal}. Black arrow indicates $J_{13}=0$. Simulation parameters: the nondiagonal and nonzero elements of $\mathbb{W}$ are $W_{21}=10.8$, $W_{31}=13.4\me^{-B_{13}-F_{13}/2}$, $W_{41}=16.2$, $W_{12}=94.8$, $W_{32}=26.6$, $W_{13}=45.5 \me^{-B_{13}+F_{13}/2}$, $W_{23}=19.5$, $W_{43}=14.3$, $W_{14}=0.5$, $W_{34}=9.8$, where $B_{13}=1$. The forces in the inset are $F_1=F_{13}-0.6$, $F_2=F_{13}+4.4$ which give $|F_1|=|F_2|$ at $F_{13}=-1.9$.
   }
    \label{fig:exact}
\end{figure}

In \cref{fig:exact}, we consider the responses $\partial_{B_{13}}\pi_i$ with $i=1,\dots,4$, for the network given in the inset, and compare it to the bound $|\partial_{B_{nm}}\pi_i|\leq \pi_i(1-\pi_i)\tanh(F_\text{max}/4)$ obtained in \cite{owen2020universal}. We see that $J_{13}$ can vanish even at nonzero value $F_\text{max}=\text{max}(|F_1|,|F_2|)\neq 0$. In other words, the system is out-of-equilibrium but the edge $1-3$ is detailed balanced.

\section{Numerical simulations} \label{sec:simulations}

\begin{figure}[h]
    \centering
    \includegraphics[width=0.45\textwidth]{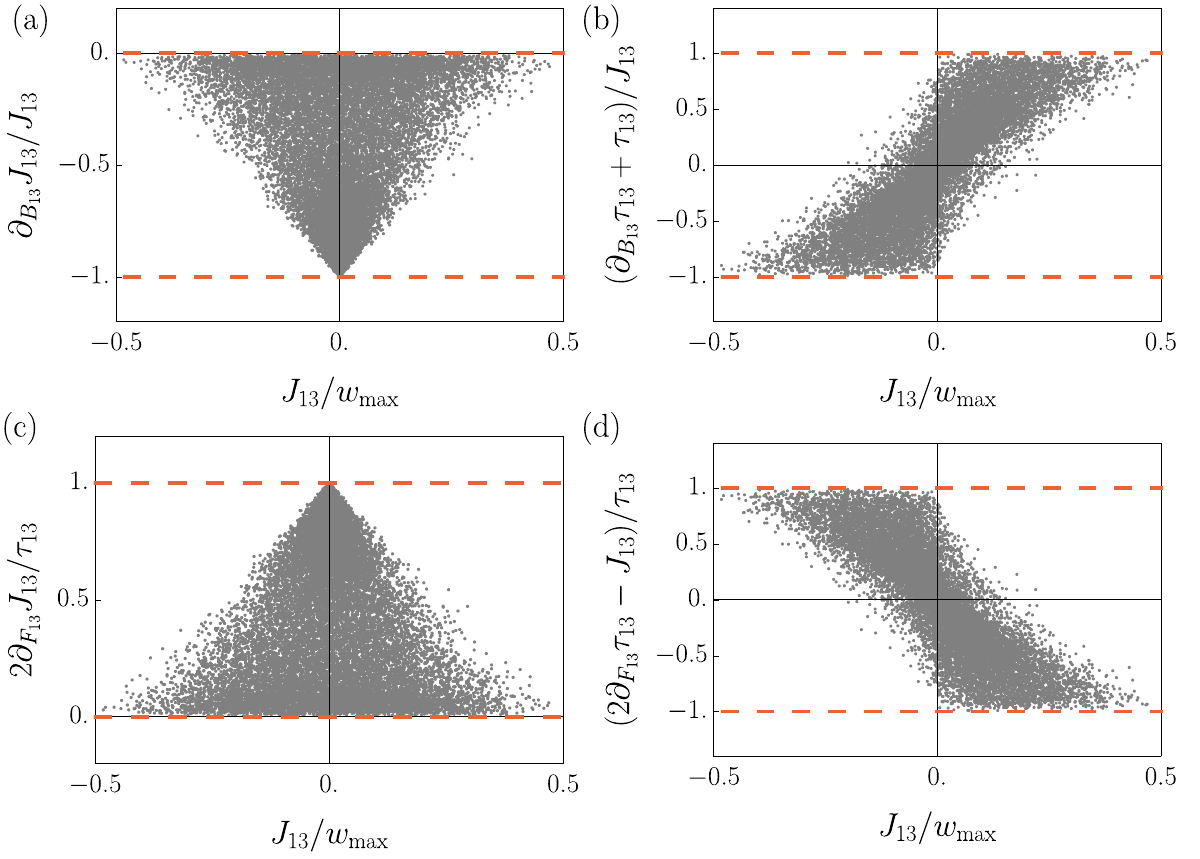}
    \caption{a-d: Illustrations of the bounds of $J_{nm}$ from \cref{tab:main-results} and $\tau_{nm}$ from \cref{eq:tau-tight-bounds}. The dashed lines show the corresponding bounds. The dots are the result of numerical calculations for 20000 random matrices $\mathbb W$ with a homogeneous distribution of elements in the range $(0,w_\text{max})$. The network corresponds to the inset in \cref{fig:exact}, $w_\text{max}=100$.}
    \label{fig:random}
\end{figure}

In \cref{fig:random}, we numerically verify the bounds in \cref{tab:main-results} and \cref{eq:tau-tight-bounds} using random generated rate matrices for the network shown in the inset of \cref{fig:exact}. 

\newpage

\bibliography{biblio}
\end{document}